\documentclass[
superscriptaddress,twocolumn,
amsmath,amssymb,
longbibliography
]{revtex4-2}

\usepackage{graphicx}
\usepackage{dcolumn}
\usepackage{bm}
\usepackage{braket}
\usepackage{comment}
\usepackage{xcolor}
\usepackage{soul}

\usepackage{hyperref}
\hypersetup{colorlinks=true,linkcolor=blue,citecolor=blue,urlcolor=blue}

\newcommand{\tr}[1]{\mathrm{Tr}\left( #1 \right)}

\DeclareMathOperator{\var}{var}
\def\Re{\operatorname{Re}}
\def\Im{\operatorname{Im}}

\begin{document}

\title{Interferometry of quantum correlation functions to access quasiprobability distribution of work}

\author{Santiago Hern\'{a}ndez-G\'{o}mez}
\email{shergom@mit.edu,hernandez@lens.unifi.it}
\affiliation{Research Laboratory of Electronics, Massachusetts Institute of Technology, Cambridge, MA 02139, USA}

\author{Takuya Isogawa}
\affiliation{Research Laboratory of Electronics, Massachusetts Institute of Technology, Cambridge, MA 02139, USA}
\affiliation{Department of Nuclear Science and Engineering, Massachusetts Institute of Technology, Cambridge, MA 02139, USA}

\author{Alessio Belenchia}
\affiliation{Institute of Quantum Technologies, German Aerospace Center (DLR), D-89077 Ulm, Germany}

\author{Amikam Levy}
\affiliation{Department of Chemistry, Institute of Nanotechnology and Advanced Materials and the Center for Quantum
Entanglement Science and Technology, Bar-Ilan University, Ramat-Gan, 52900 Israel}

\author{Nicole Fabbri}
\affiliation{Istituto Nazionale di Ottica del Consiglio Nazionale delle Ricerche (CNR-INO), I-50019 Sesto Fiorentino, Italy}
\affiliation{European Laboratory for Non-linear Spectroscopy (LENS), Universit\`a di Firenze, I-50019 Sesto Fiorentino, Italy}

\author{Stefano Gherardini}
\affiliation{European Laboratory for Non-linear Spectroscopy (LENS), Universit\`a di Firenze, I-50019 Sesto Fiorentino, Italy}
\affiliation{Istituto Nazionale di Ottica del Consiglio Nazionale delle Ricerche (CNR-INO), Largo Enrico Fermi 6, I-50125 Firenze, Italy}
\affiliation{SISSA, via Bonomea 265, I-34136 Trieste, Italy}

\author{Paola Cappellaro}
\affiliation{Research Laboratory of Electronics, Massachusetts Institute of Technology, Cambridge, MA 02139, USA}
\affiliation{Department of Nuclear Science and Engineering, Massachusetts Institute of Technology, Cambridge, MA 02139, USA}


\begin{abstract} 
The Kirkwood-Dirac quasiprobability distribution, intimately connected with the quantum correlation function of two observables measured at distinct times, is becoming increasingly relevant for fundamental physics and quantum technologies. 
This quasiprobability distribution can take non-positive values, and its experimental reconstruction becomes challenging when expectation values of incompatible observables are involved. 
Here, we use an interferometric scheme aided by an auxiliary system to reconstruct the Kirkwood-Dirac quasiprobability distribution. 
We experimentally demonstrate this scheme in an electron-nuclear spin system associated with a nitrogen-vacancy center in diamond. 
By measuring the characteristic function, we reconstruct the quasiprobability distribution of work and analyze the behavior of its first and second moments. 
Our results clarify the physical meaning of the work quasiprobability distribution in the context of quantum thermodynamics. 
Finally, we study the uncertainty of measuring the Hamiltonian of the system at two times, via the Robertson-Schr{\"o}dinger uncertainty relation, for different initial states.
\end{abstract}


\maketitle

\section{Introduction}
While probabilities intrinsically emerge from measurements in quantum theory, observable incompatibility precludes a general state of a quantum mechanical system from being represented in terms of joint probabilities over its phase space. 
Fundamental to the celebrated Heisenberg uncertainty principle~\cite{heisenberg1985anschaulichen,busch2014heisenberg} and the information-disturbance trade-offs of quantum measurements~\cite{busch2013proof,branciard2013error}, non-commuting observables also prevent the representation of quantum states and processes as a joint probability distribution over their measurement outcomes.
Nonetheless, it was recognized early on that states can be represented in terms of \textit{quasiprobabilities}, i.e., joint distributions satisfying all but one of Kolmogorov axioms: they can take non-positive values~\cite{cohen1995time,wigner1997quantum}. 
In Ref.~\cite{SpekkensPRL2008}, it is shown that the negativity of all quasiprobability-distribution representations of the experiment under scrutiny is related to non-classicality as witnessed by contextuality~\cite{HofmannPRL2012,PuseyPRL2014,Hofer2017quasi}. This general result linking negativity and quantum contextuality can be linked to the {\it Kirkwood-Dirac quasiprobability} (KDQ)~\cite{lostaglio2018quantum,ArvidssonShukur2024review}, introduced by Kirkwood~\cite{kirkwood1933quantum} and Dirac~\cite{dirac1945analogy} in the first half of the twentieth century as a representation of the quantum state over incompatible observables. 
In recent years, there has been a renewed interest in the KDQ~\cite{yunger2018quasiprobability,DeBievrePRL2021,companion_theory_paper,SantiniPRB2023,BudiyonoPRAquantifying,FrancicaPRE2023,wagner2023quantum,ArvidssonShukur2024review},  
especially on the study of its non-reality and non-positivity, and its link 
to quantum metrological advantages in both local and postselected scenarios~\cite{lupu2021negative}, to  weak values~\cite{aharonov1988how,Dressel2014colloquium,KunjwalPRA2019}, and to benefits in quantum thermodynamics~\cite{diaz2020quantum,levy2020quasiprobability,maffei2022anomalous,hernandez2022experimental,PeiPRE2023,GherardiniTutorial,Upadhyaya2023arxiv}. Moreover, due to its inherent link with quantum correlation functions, it is quickly finding applications in several fields of physics, from condensed matter to quantum chaos~\cite{SilvaPRL2008,yunger2018quasiprobability,dressel2018strengthening,chenu2018quantum,mohseninia2019strongly,alonso2019out}.

To define the KDQ, we introduce two, generally time-dependent, observables $A=\sum_{j}a_{j}\Pi_{j}^{A}$ and $B=\sum_{k}b_{k}\Pi_{k}^{B}$, written in terms of their spectral decomposition, where $\Pi^C_{\ell}(t) = \left|C_{\ell}(t)\right\rangle\!\!\left\langle C_{\ell}(t)\right|$ are the eigenprojectors of a Hermitian operator $C$. 
Then, the KDQ for a closed quantum system is defined as
\begin{equation}\label{eq:bare_KDQ}
q_{jk}(\rho) = \tr{ U^\dag \Pi_k^B \, U\Pi_j^A \, \rho },
\end{equation}
where $U$ is the unitary operator describing the time evolution of the system and $\rho$ is its initial density operator. 
Notice that the KDQ can be generally defined for any pair of projectors, here we opt to use the eigenprojectors $\Pi_j^A$ and $\widetilde{\Pi}_k^B \equiv U^\dagger \Pi_k^B U$ such that the KDQ 
characterizes the two-time quantum correlation function of incompatible observables given an initial state that evolves under $U$~\cite{margenau1961correlation,companion_theory_paper}.

While the KDQ encodes a full description of the quantum state and correlation between incompatible observables, accessing it is not straightforward. 
The assessment of a quasiprobability distribution has been the subject of intense investigation, both theoretically and experimentally~\cite{johansen2007quantum,Buscemi_2014,Buscemi2013direct,Solinas2016probing,SolinasPRAmeasurement,companion_theory_paper,SolinasPRA2022,hernandez2022experimental,wagner2023quantum,GherardiniTutorial}. 
The traditional method of sequentially performing projective measurements on two observables, also referred to as the two-point measurement (TPM) scheme within the field of quantum thermodynamics~\cite{campisi2011colloquium}, is inadequate. This is because the initial measurement eliminates the quantum coherence and correlations in the initial state and alters the statistics of all non-commuting observables~\cite{HofmannNJP2014}.
The \textit{real} part of the KDQ could be reconstructed by combining the results of a series of projective measurements, including the ones from the TPM scheme~\cite{johansen2007quantum,hernandez2022experimental}. However, the imaginary part of the KDQ cannot be obtained in this way. 
The close relation between the KDQ and weak values~\cite{companion_theory_paper} implies that the KDQ could be obtained using schemes originally developed for direct wave-function measurements or for state tomography, which were either based on weak measurements~\cite{hofmann2010complete,lundeen2011direct,PhysRevLett.108.070402,PhysRevLett.92.130402,PhysRevLett.102.020404}, or on more refined projective measurement schemes~\cite{bamber2014observing,thekkadath2016direct,piacentini2016measuring,kim2018direct,calderaro2018direct}. 
Here we opt for a more direct approach, where the characteristic function of the KDQ is the output of the experimental scheme. This approach is more general, as it avoids the requirement of weak coupling imposed by weak measurement schemes. More interestingly, moving away from state tomography-based schemes allows us to focus on the system dynamics, that is, reconstructing the KDQ for observables at two different times,  exploring the effects of the system evolution, and investigating in particular its thermodynamic properties.

In this work, we experimentally reconstruct the full KDQ via an interferometric scheme adapted to the nitrogen-vacancy (NV) center in diamond~\cite{Hirose2016, Chen2022}. The system of interest in our scheme is represented by the electronic spin in the NV center, while the nitrogen nuclear spin acts as an auxiliary system. Ramsey-type interferometry~\cite{mazzola2013measuring,mazzola2014detecting,BatalhaoPRL14} allows the reconstruction of the characteristic function of a KDQ distribution, for a generic initial state of the system. 
While the scheme we employ is quite general and valid for any choice of observables, in the experimental realization, we focus on the KDQ characterizing the two-time quantum correlation function of the (initial and final) energy of the system, namely the {\it energy autocorrelation function}. In turn, this allows us to interpret our result in a thermodynamic framework, connecting with previous results on work fluctuations and showing how quantum coherence can be the sole source to attain extractable work in the system.
Furthermore, the imaginary part of the KDQ --- that we obtain via our interferometric reconstruction --- is directly connected to the Robertson-Schr{\"o}dinger uncertainty relation~\cite{RobertsonPR1929,SchrodingerPmK1930,schrodinger1999heisenberg}. This approach enables us to quantify the uncertainty in recording the initial and final energies of the system in our experiment by measuring their commutator.

\section{Results}
\subsection{Interferometric scheme for KDQ}

\begin{figure}
    \centering
    \includegraphics[width=\columnwidth]{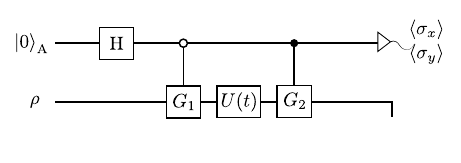}
    \caption{
    Interferometric scheme quantum circuit to obtain the characteristic function of KDQ distributions of work. The KDQ characteristic function is codified into the coherence of an auxiliary qubit, which is readout by means of a Ramsey scheme. Here, $\mathrm{H}$ is the Hadamard gate, $\sigma_{x,y}$ are the Pauli matrices, $U(t)$ is the unitary operator describing the system dynamics, and the gates $G_{1,2}$ are defined in the text.
    }
    \label{fig:scheme_original}
\end{figure}

The evolution of a closed quantum system under a work protocol is described by the unitary operator $U(t) =  \mathcal{T} \exp(- i\int_0^t H(t') \, {\rm d}t' )$, where $H(t)$ is the time-varying Hamiltonian of the system. Here and throughout this text we take $\hbar=1$.  
Let us write the Hamiltonian in terms of its spectral decomposition $H(t) = \sum_{j=1}^N E_j(t) \Pi_j(t)$, where $\Pi_j(t) = \left|E_j(t)\right\rangle\!\!\left\langle E_j(t)\right|$ is the eigenprojector of the Hamiltonian and $N$ is the dimension of the system. The KDQ work distribution~\cite{companion_theory_paper,GherardiniTutorial} for this work protocol is therefore defined in terms of the energy eigenprojectors $\Pi_f(t)$ and $\Pi_i(0)$:
\begin{equation}\label{eq:kd}
q_{if}(\rho) = \tr{ U^\dag(t) \Pi_f(t) U(t) \; \Pi_i (0) \, \rho },
\end{equation}
where $\rho$ is an arbitrary initial state of the system.

The quasiprobability distribution, Eq.~(\ref{eq:kd}), reduces to the two-point-measurement (TPM) distribution whenever $\Pi_i(0)$ commutes with either $\rho$ or $U^\dag(t) \Pi_f(t) U(t)$~\cite{companion_theory_paper}:
\begin{equation}\label{eq:classical_limit_qthermo}
q_{if}(\rho) = {\rm Tr}\left(\Pi_i(0)\rho\right)\,{\rm Tr}\left(  U^\dag(t) \Pi_f(t) U(t) \Pi_i(0)\right).
\end{equation} 
Effective commutativity can result from decoherence or coarse-graining of the energy measurements. In this limit, the KDQ coincides with the statistics originating from the TPM scheme which, in turn, can be associated to the classical definitions of work in the case of unitary dynamics~\cite{JarzynskiPRX2015}\footnote{Note that, as outlined in Refs.~\cite{ArvidssonShukurJPA2021,DeBievrePRL2021,DeBievreArXiv2022}, a positive KDQ distribution can be attained even in the non-commutative case where $[\Pi_i (0),\rho]\neq 0$, $[U^\dag(t) \Pi_f(t) U(t),\rho]\neq 0$, and 
$[H(0),H(t)]\neq 0$ for some $i,f,t$, albeit the KDQ distribution still remains different from the one of the TPM scheme.}. 
In general, the KDQ is complex valued and its real part can be negative. Thus, it cannot be accessed by means of a procedure based on sequential projective measurements (the TPM scheme).

\begin{figure*}
    \centering
    \includegraphics[width=\textwidth]{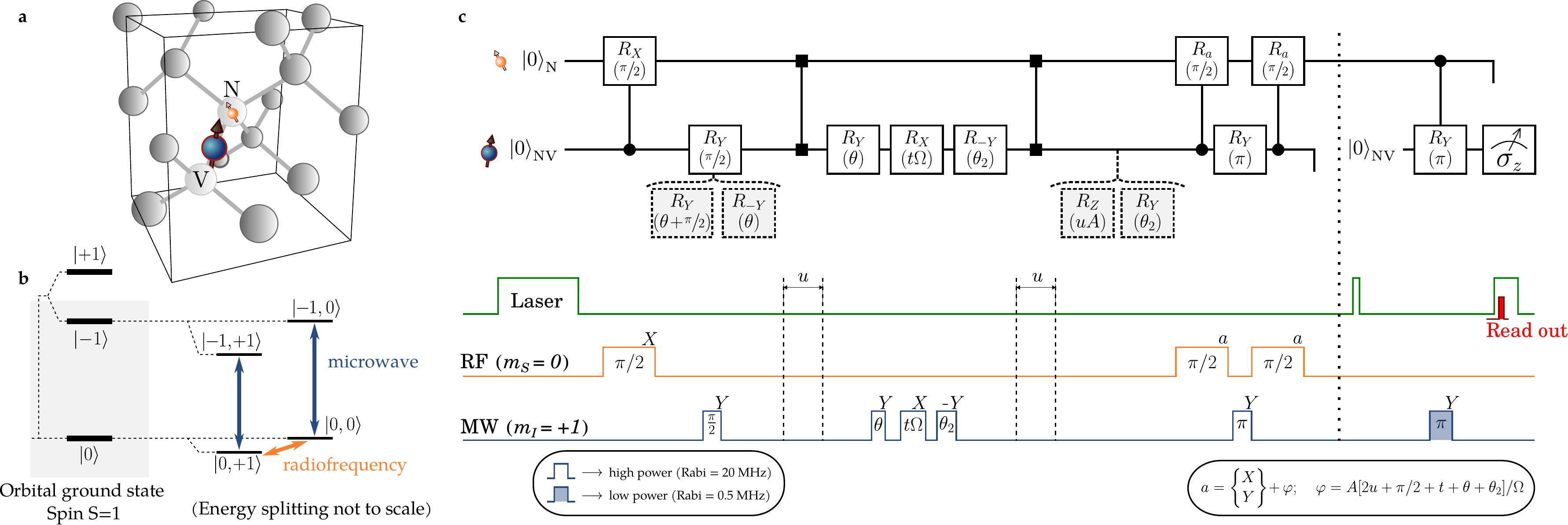}
    \caption{\textbf{a} NV center in the diamond lattice. Blue indicates the electronic spin, and orange denotes the nitrogen nuclear spin. \textbf{b} Energy levels that form the two-qubit system. The NV electronic (nitrogen nuclear) spin transitions can be coherently driven by applying resonant microwave (radiofrequency) pulses. The notation is $\ket{m_S,m_I}$.
    \textbf{c} Adaptation of the interferometric scheme shown in Fig.~\ref{fig:scheme_original} for our experimental platform: NV center electronic spin and nitrogen nuclear spin. On the top: quantum circuit; on the bottom: pulse sequence. 
    The vertical lines with black squares correspond to the free evolution during a time $u$. 
    The quantum gates are rotations along X, Y, or a. When $a = \varphi $ ($a = \pi/2 + \varphi $) the final measurement corresponds to $\langle\sigma_x\rangle$ ($\langle\sigma_y\rangle$), where $\varphi \equiv A[2u + \pi/2 + t+\theta + \theta_2]/\Omega$ is the phase acquired by the nuclear spin during the Ramsey scheme due to the hyperfine term of the Hamiltonian. The `duration' of each quantum gate is indicated in parentheses. 
    Notice that, the gates $R_Y(\theta +\pi/2)$ (used to prepare the initial state $\rho$) and $R_{-Y}(\theta)$ (part of the gate $G_1$) can be simplified into $R_Y(\pi/2)$. Moreover, the gates $R_Z(uA)$ and $R_{Y}(\theta_2)$ can be ignored since they do not affect the state of the nuclear spin before the readout. 
    Further details about the scheme can be found in Methods.}
    \label{fig:NV_lattice_and_E_levels_scheme_experiment}
\end{figure*}

In this work, we aim to obtain both the real and imaginary parts of KDQ. For this purpose, we experimentally reconstruct the characteristic function
\begin{align}\label{eq:char_func}
\mathcal{G}(u) &\equiv \int e^{ iuW} P(W) \mathrm{d}W \notag \\ 
&= \sum_{i,f}q_{if}(\rho)\; e^{ iu (E_f -E_i)} \notag\\
&= {\rm Tr}\left[ e^{-i u H(0)} \rho \, U^\dagger e^{i u H(t)} U \right]  
\end{align} 
associated with the KDQ distribution of energy differences $E_f-E_i$ that represent the realization of the work ($W$) random variable 
\begin{equation} \label{eq:PWandqif}
    P(W) = \sum_{i,f}q_{if}(\rho)\;\delta\left( W-(E_f-E_i) \right),
\end{equation}
where $\delta(\cdot)$ denotes the Dirac delta function. 
To do so, we implement the quantum circuit shown in Fig.~\ref{fig:scheme_original}. 
This circuit is essentially the same as the ones in Refs.~\cite{mazzola2013measuring,dorner2013extracting,BatalhaoPRL14}, with the important difference that we do not assume that the initial state of the system $\rho$ is a mixed thermal state~\cite{companion_theory_paper,GherardiniTutorial}. 
Here we will show that, as theorized in Ref.~\cite{companion_theory_paper}, this 
scheme can be used to reconstruct the whole KDQ characteristic function for non-thermal initial states.

The key idea of the interferometric scheme is to perform a Ramsey scheme on the auxiliary qubit. During the \emph{free} evolution of the Ramsey scheme, the auxiliary qubit is put in contact with the quantum system by means of two conditional gates $G_1$ and $G_2$, which are defined as:
\begin{align}
G_1(u) &= e^{-i u H_0} \otimes \ket{0}_{A}\!\!\bra{0} +  \mathbb{I} \otimes \ket{1}_{A}\!\!\bra{1} \label{eq:gate_G1}\\
G_2(u) &= \mathbb{I} \otimes \ket{0}_{A}\!\!\bra{0} + e^{-i u H_t} \otimes \ket{1}_{A}\!\!\bra{1}, \label{eq:gate_G2}
\end{align}
where $u$ is the gate duration. Note that we have introduced the notation $H_t=H(t)$ to explicitly indicate that the Hamiltonian is not changing during the duration $u$. 
In between these two gates, the system evolves under its Hamiltonian for a time $t$ as described by the unitary operator $U(t)$. 
Finally, the real and imaginary parts of the characteristic function [Eq.~\eqref{eq:char_func}] are encoded in the expectation values of the auxiliary qubit Pauli matrices as $\left\langle \sigma_x\right\rangle_{\rm A} = {\rm Re} \left[\mathcal{G}(u) \right]$ and $\left\langle \sigma_y\right\rangle_{\rm A} = {\rm Im} \left[\mathcal{G}(u) \right]$ respectively.

\subsection{Experiment}
\label{sec:experiment}

\begin{figure*}
    \centering
    \includegraphics[width=0.9\textwidth]{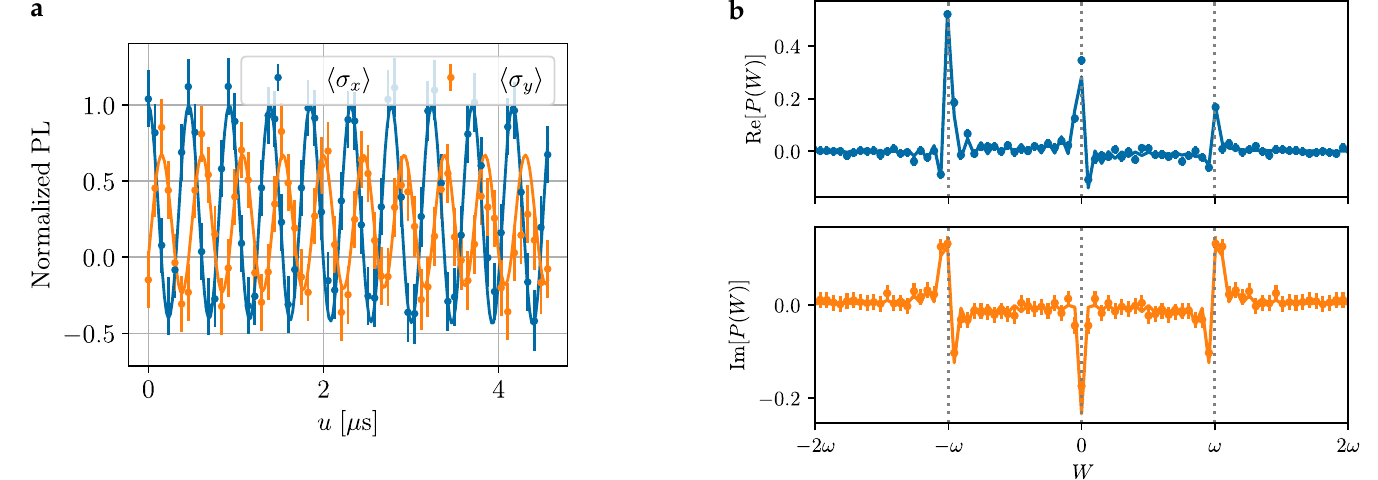}
    \caption{
    Example of the experimental results for a single value of time $t=26$~ns $=7\pi/6\Omega$. 
    \textbf{a} Characteristic function $\mathcal{G}(u)$~[Eq.~\eqref{eq:char_func}] of the KDQ distribution of work. The real~(imaginary) part of the characteristic function, shown in blue~(orange), is the result of measuring the expectation value of $\sigma_{x(y)}$ with respect to the state of the auxiliary qubit in the interferometric scheme. The expectation value $\langle\sigma_{x(y)}\rangle$ is the photoluminescence~(PL) intensity normalized with respect to the reference intensity of the eigenstates of the observable $\sigma_{x(y)}$ .
    \textbf{b} By applying a Fourier transform to $\mathcal{G}(u)$, we are able to reconstruct the KDQ distribution $P(W)$ of work. The error bars are the standard deviation of the FFT calculated as the mean (over all the data) standard deviation of $\mathcal{G}(u)$. The values we are interested in are those where $W=\pm\omega,0$ (indicated with vertical dashed lines), which are the only allowed energy variations for the two-level system evolving under the Hamiltonian $H(t)$~[Eq.~\eqref{eq:H_t}]. Note that the real and imaginary parts of $\mathcal{G}(u)$ are both necessary to obtain the real or imaginary parts of the work distribution $P(W)$.}
    \label{fig:charFunc_and_IFFT}
\end{figure*}

In this section, we show how to implement the interferometric scheme using the electronic and nuclear spins of a single NV center. 
We consider the case where the spin qubit  evolves under the Hamiltonian
\begin{equation}
H(t) = \frac{1}{2}\left[ \Omega\left( \cos(\delta t)\,\sigma_x + \sin(\delta t)\,\sigma_y\right)  +\delta \sigma_z \right] \label{eq:H_t},
\end{equation}
such that the energy eigenstates $\ket{E_j(t)}$ are time dependent, but its eigenvalues are $\pm\omega/2\equiv\pm\sqrt{\Omega^2+\delta^2}/2$ for every time $t$.
In our experiments, we set the parameters in Eq.~\eqref{eq:H_t} to $\delta = \Omega \sqrt{3}$ and $\Omega=2\pi\,875/39$~MHz. 
We will use the NV electronic spin qubit as the system that evolves under the Hamiltonian $H(t)$~[Eq.~\eqref{eq:H_t}] starting from an initial pure coherent state, 
$\rho = \ket{+}\!\!\bra{+}$, with $\ket{+} \equiv (\ket{E_0(0)} + \ket{E_1(0)})/\sqrt{2}$ a superposition of the eigenstates of $H(0)$. The Hamiltonian in Eq.~\eqref{eq:H_t}, while quite general, provides a convenient implementation since the corresponding evolution unitary operator can be written as $U(t)= \mathcal{T} \exp(- i\int_0^t H(t') \, {\rm d}t' )=\exp\left(-i  t \delta \sigma_z /2 \right)\exp\left(-i  t \Omega \sigma_x /2 \right) \otimes \mathbb{I}$. In turn, this allows to simplify the original circuit (Fig.~\ref{fig:scheme_original}), 
replacing the evolution with the simpler operator $U_{B}(t) = e^{-i  t \Omega \sigma_x /2} \otimes \mathbb{I}$ and the gate $G_2(u)$~[Eq.~\eqref{eq:gate_G2}] 
with 
\begin{equation}
G_B(u) = \mathbb{I} \otimes \ket{0}_{\rm A}\!\!\bra{0} + e^{-i u H_0}  \otimes \ket{1}_{\rm A}\!\!\bra{1},  \label{eq:G_B}
\end{equation}
while obtaining the same real and imaginary parts of the characteristic function of the work KDQ distribution~(see Supplementary Note 1 for more details).

The details on how to implement such interferometric scheme are summarized in Fig.~\ref{fig:NV_lattice_and_E_levels_scheme_experiment}. 
There, the system $\rho$ corresponds to the electronic spin $\vec{S}$, and the auxiliary system is the nuclear spin $\vec{I}$ of the  nitrogen, as depicted in Fig.~\ref{fig:NV_lattice_and_E_levels_scheme_experiment}a,b. 
Both spins form triplets, and their total Hamiltonian reads $H_{SI} = \Delta S_z^2 + \gamma_e \vec{B}(t)\cdot\vec{S} + Q I_z^2 + \gamma_n \vec{B}(t)\cdot\vec{I} + A S_z I_z $, where $\Delta\simeq 2.87$~GHz is the zero-field-splitting, $Q = -4.95$~\cite{Manson06,Smeltzer09} is the nuclear quadrupole moment, $A\simeq -2.16$~MHz is the hyperfine coupling constant~\cite{Smeltzer09}, $\gamma_e=2.8$MHz/G and $\gamma_n=-0.308$ kHz/G are the electron and nuclear gyromagnetic ratios, and $\vec{B}(t)$ is a magnetic field. A static bias magnetic field aligned with the NV quantization axis $B_z = 357.07 \pm 0.13$ G splits the $m_S=\pm1$ levels. 
In addition, an AC field $B_x(t) = B \cos(2\pi\nu t + \phi)$ is used to drive either the electronic or the nuclear spins. For our experiments, we need two qubits, so we ignore the $m_S=+1$ and the $m_I=-1$ levels to obtain a two-qubit system. 
The system is optically initialized into the state $\left|m_S=0,m_I=+1\right\rangle \equiv \ket{0,1}_{\textrm{qubit}}$ at the beginning of each experiment. 
Therefore, the Hamiltonian $H_{SI}$ can be reduced to a two-qubit system Hamiltonian. In the frame rotating  at the resonant frequency of the electronic spin ($\omega_e \equiv \Delta - \gamma_e B_z$) and  the nuclear spin ($\omega_n \equiv Q + \gamma_n B_z$), using the rotating wave approximation, we can write
\begin{align}\label{eq:Hamiltonian_e_n}
H_{\rm tot} = & H_e \otimes \mathbb{I}  + \mathbb{I} \otimes H_n  + H_{\rm I} \,. 
\end{align}
In \eqref{eq:Hamiltonian_e_n} $H_{\rm I}$ is the interaction Hamiltonian, i.e.,
\begin{align}
    H_{\rm I} &= \frac{A}{4}(\sigma_z\otimes \mathbb{I} + \mathbb{I} \otimes \sigma_z - \sigma_z\otimes\sigma_z), \label{eq:H_I}
\end{align}
and $H_{e(n)}$ describe the electronic and nuclear spin (control) Hamiltonians, given by 
\begin{align} 
    H_{e(n)} &= \frac{\Omega_{e(n)} }{2} \left( \cos(\phi)\,\sigma_x + \sin(\phi)\,\sigma_y \right)\,, \label{eq:H_en}
\end{align}
where $\Omega_{e(n)} = \gamma_{e(n)} B/\sqrt{2}$ is the Rabi frequency for the electronic~(nuclear) spin, and $\phi$ is the phase of the AC field. 
The control field applied to the nuclear spin can only achieve selective rotations, because the condition $\Omega_n\ll A$ always holds, due to the relatively small gyromagnetic ratio of the nuclear spin. 
In contrast, we can apply non-selective pulses with high microwave power ($\Omega_e\gg A$)  to the electronic spin. 
One could think that these control fields would be enough to implement the scheme in Fig.~\ref{fig:scheme_original}. However, the presence of the hyperfine term $H_\mathrm{I}$ would affect the expectation values of $\sigma_{x,y}$ for the ancilla spin, hence defeating the purpose of the interferometric scheme. Instead, the combination of these nuclear and electronic spin gates is sufficient to implement the quantum circuit shown in Fig.~\ref{fig:NV_lattice_and_E_levels_scheme_experiment}c.  By choosing the parameters $\Omega_{e}=\Omega$, $\theta = \arctan(\Omega/\delta)$, and $\theta_2 = \theta + \pi$, we ensure that the real and imaginary parts of the characteristic function measured with the circuit in Fig.~\ref{fig:NV_lattice_and_E_levels_scheme_experiment}c are the same as the ones measured with the circuit in Fig.~\ref{fig:scheme_original} and are therefore associated to the work protocol determined by the system Hamiltonian $H(t)$ [Eq.~\eqref{eq:H_t}]. 
Additional details on the implementation of this circuit in our platform are given in the Methods section. 

\begin{figure}
    \centering
    \includegraphics[width=\columnwidth]{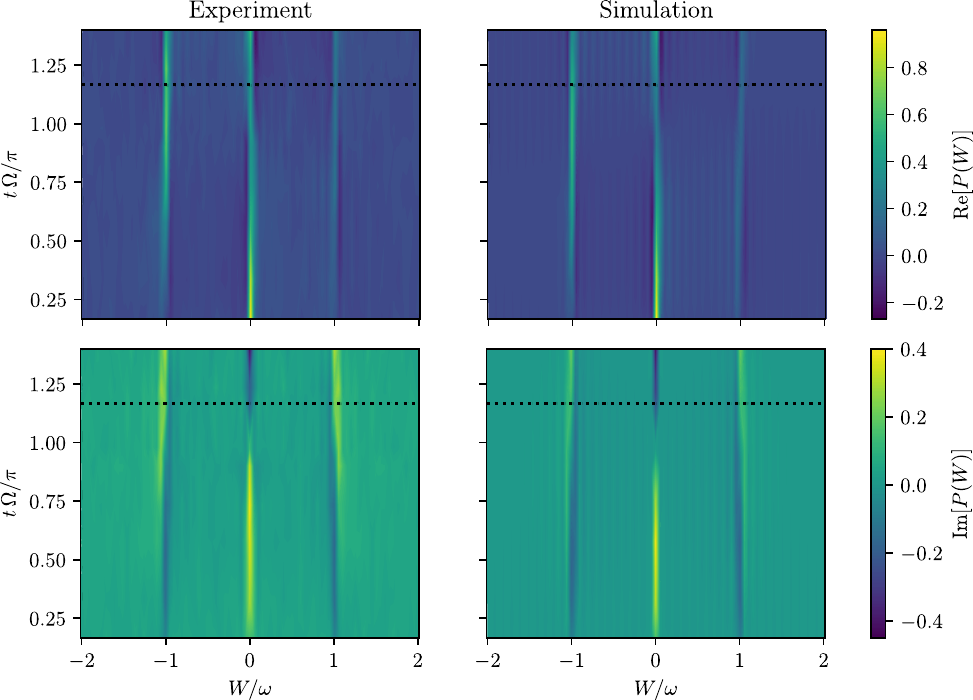}
    \caption{KDQ distribution $P(W)$ of work for different values of the evolution time $t$. The horizontal dashed lines correspond to the data shown in Fig.~\ref{fig:charFunc_and_IFFT}(b).
    Top~(bottom) panels show the real~(imaginary) part of $P(W)$. The left panels indicate the experimental results, obtained by doing a Fourier analysis of the measured characteristic function. The right panels denote the distribution obtained by simulating the KDQ $q_{if}$ for a qubit that evolves under $H(t)$~[Eq.~\eqref{eq:H_t}], then calculating the characteristic function, and finally applying a Fourier analysis. There are no fitted parameters between experiments and simulations.}
    \label{fig:IFFTdata}
\end{figure}

\begin{figure*}
    \centering
    \includegraphics[width=\textwidth]{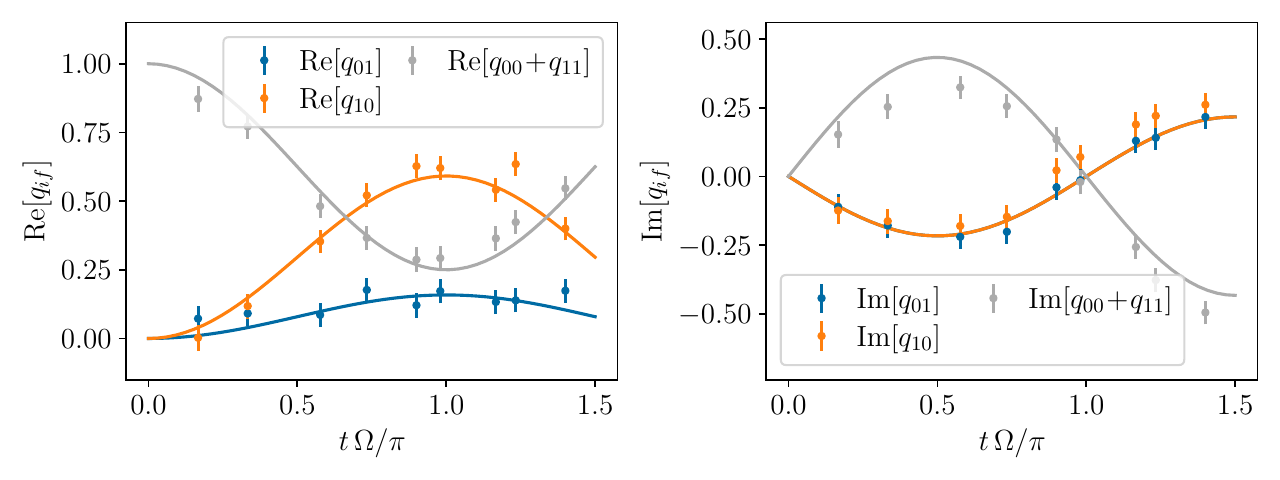}
    \caption{
    Reconstruction of the real and imaginary values of the KDQ $q_{if}$. The experimental values (markers with errorbars) are obtained by taking the work distribution in Fig.~\ref{fig:IFFTdata}~(left panels) and considering only the values around $W = \pm \omega,0$. 
    In contrast, the solid line indicates values of $q_{if}$ simulated using Eq.~\eqref{eq:kd} for a qubit evolving under the Hamiltonian $H(t)$.  
    The case with no net energy change corresponds to $q_{00}$ and $q_{11}$. The system energy increases~(decreases) by a factor $\omega$~($-\omega$) for $q_{01}$~($q_{10}$). Notice that the real part of the KDQ, also known as Margenau-Hill quasiprobability MHQ, is always positive. However, the imaginary part of the KDQ is different from zero. Therefore, in the present case study, measuring the MHQs is not enough to witness the non-classicality entailed by non-commutativity. For that, it is necessary to measure the full KDQs.}
    \label{fig:KDQ_re_and_im}
\end{figure*}

An example of the data obtained with the interferometric scheme for a single value of the time $t$ while varying the gate duration $u$ is shown in Fig.~\ref{fig:charFunc_and_IFFT}(a). 
The expectation values $\left\langle \sigma_x\right\rangle_{\rm A} = {\rm Re} \left[\mathcal{G}(u) \right]$ and $\left\langle \sigma_y\right\rangle_{\rm A} = {\rm Im} \left[\mathcal{G}(u) \right]$ with respect to the state of the auxiliary qubit yield the real and imaginary part of the characteristic function, which we use to reconstruct the work distribution. 

\subsection{KDQ work distribution}

We obtain the work distribution shown in Fig.~\ref{fig:charFunc_and_IFFT}(b) by performing a fast Fourier transform (FFT) to the characteristic function $\mathcal{G}(u)$ [see Eq.~\eqref{eq:char_func}]. 
We have repeated the same set of measurements but for different values of the time $t$. The results for all the experiments are shown in Fig.~\ref{fig:IFFTdata}, left panels. 
These should be compared with the numerical simulations in the right panels of Fig.~\ref{fig:IFFTdata}, which were derived by calculating the characteristic function of the KDQ using Eq.~\eqref{eq:kd}, for a two-level system evolving under the Hamiltonian in Eq.~\eqref{eq:H_t}. 
It is worth pointing out the remarkable agreement between simulation and experiment, especially considering that there are no fitted free parameters. 
To quantitatively compare data and simulation in Fig.~\ref{fig:IFFTdata}, we have calculated the reduced chi squared~\footnote{The reduced chi squared was obtained as: $\chi^2_n = \frac{1}{n} \sum_{i}^n (x_i - s_i)^2/\sigma_i^2$, where $n$, $x_i$, $\sigma_i$ and $s_i$ represent respectively the number of data points, the experimental data, its standard deviation and the simulated data.} by obtaining $\chi^2_n\left({\rm Re} \, P(W)\right) = 1.51$ and $\chi^2_n\left({\rm Im} \, P(W)\right) = 1.06$. Hence, the agreement between simulation and experiment is, on average, comparable with the experimental precision.

Having reconstructed the distribution $P(W)$, it is then possible to extract the KDQ itself from the experimental data shown in Fig.~\ref{fig:IFFTdata}(a). 
The experimental reconstruction of the $q_{if}$ values could be achieved by considering only the points at $W=E_f(t)-E_i(0)$, see Eq.~\eqref{eq:PWandqif}. In this experimental study, the only allowed values of $E_f(t)-E_i(0)$ are $\pm \omega, 0$, see Eq.~\eqref{eq:H_t}. 
However, to reduce the experimental errors, it is better to compute the average of $P(W)$ over a small interval around those values. Here, we consider an interval covered by seven experimental points centered around $W=\pm \omega$ and $W=0$, and we integrate them to obtain each $q_{if}$. This interval was chosen considering that, on average for all the data, the signal-to-noise ratio was larger than $1$ in that interval. 
The peaks around $W=+\omega$ and $W=-\omega$ correspond to $q_{01}$ and $q_{10}$, respectively. The peak around $W=0$ correspond to the sum $q_{00} + q_{11}$. 
In our case study, we are unable to individually determine $q_{00}$ and $q_{11}$ since the Hamiltonian's eigenvalues remain constant in time. Both $q_{00}$ and $q_{11}$ are associated with an energy variation equal to zero, which are combined into the same term in the characteristic function~[Eq.~\eqref{eq:char_func}].

The reconstructed values of the KDQ are shown in Fig.~\ref{fig:KDQ_re_and_im}, which is one of the main results of the paper. As mentioned before, the simulation corresponds to the computation of $q_{if}$ just using Eqs.~\eqref{eq:kd} and \eqref{eq:H_t} and without any fitted parameters. This demonstrates that it is possible to reconstruct both the real and imaginary values of the KDQ distribution using interferometry. 
The experimental data in Fig.~\ref{fig:KDQ_re_and_im} clearly follow the trend of the simulated KDQ. Notice however that the difference between experimental data and simulation is within one standard deviation for $\sim60\%$ of the points (and $\sim90\%$ are within two standard deviations). This small but systematic discrepancy between data and theory comes from two main factors. On the one hand, the finite time $u$ implies that the Fourier analysis will not perfectly reconstruct $q_{if}$, even in an ideal case without noise. On the other hand, imperfections in the experimental readout may result in small variations of the measured signal (see Fig.~\ref{fig:charFunc_and_IFFT}a) that seem negligible, but are accentuated during the Fourier analysis. These two factors are explored in more detail in the Supplementary Note 2.

\subsection{Quantum energy correlation function}

Correlation functions are broadly used in quantum mechanics to reveal quantum interference effects between systems separated in space or time. Here we focus on the latter, by studying the correlation between incompatible observables at two different times. 

\begin{figure}
    \centering
    \includegraphics[width=\columnwidth]{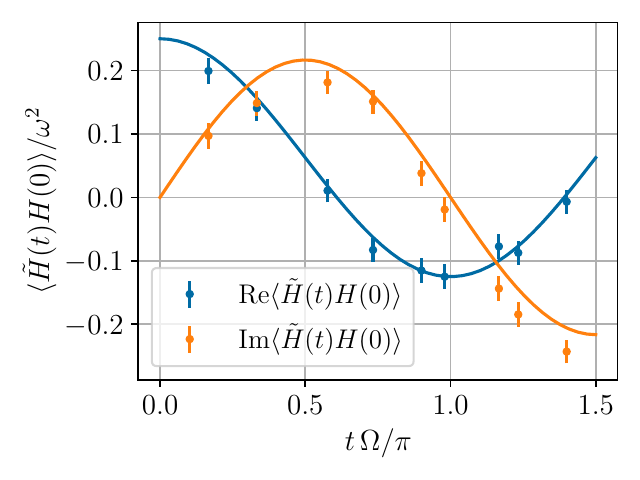}
    \caption{
    Quantum correlation function between the observables $\widetilde{H}(t) = U^\dagger H(t) U$ and $H(0)$ that are incompatible. This correlation function is directly derived from the KDQ distribution of work by using Eq.~\eqref{eq:corrFunc}. 
    The real part of the correlation function is proportional to the quantum covariance between $\widetilde{H}(t)$ and $H(0)$. Instead, the imaginary part is proportional to the expectation value of the commutator $[\widetilde{H}(t),H(0)]$, which is linked with the Robertson-Schr{\"o}dinger inequality \eqref{eq:Heisenberg_Rob_inequality} (see text). 
    Observe that, in our case, the condition $\langle \widetilde{H}(t) H(0) \rangle = 0$ is never satisfied. When the imaginary part is zero, the real part is maximized or minimized. The opposite is not true.}
    \label{fig:CorrelationFunction}
\end{figure}

The reconstruction of the KDQ allows us to extract the quantum correlation function
\begin{align}\label{eq:corrFunc}
    \langle \widetilde{H}(t) H(0) \rangle &\equiv \tr{\widetilde{H}(t) H(0) \rho} \notag \\
    &= \sum_{if} q_{if} E_i(0) E_f(t),    
\end{align}
where $\widetilde{H}(t) \equiv U^\dagger H(t) U$ is the Hamiltonian at time~$t$, evolved according to the Heisenberg picture. In Fig.~\ref{fig:CorrelationFunction} we show the real and imaginary values of the correlation function as a function of the evolution time $t$. The real part of the correlation function is
\begin{equation}
    \mathrm{Re}\langle \widetilde{H}(t) H(0) \rangle = {\rm Cov}(H(0),\widetilde{H}(t)) + \langle \widetilde{H}(t) \rangle\langle H(0) \rangle ,
    \label{eq:re_corr_func}
\end{equation} 
where $\langle\widetilde{H}(t)\rangle \equiv \tr{ \widetilde{H}(t) \rho} = \tr{H(t)\rho(t) }$, with $\rho(t) = U(t) \rho \, U^\dag(t)$, $\langle H(0)\rangle \equiv \tr{ H(0) \rho(0)}$. The covariance is as usual defined as 
\begin{eqnarray}
&{\rm Cov} (H(0),\widetilde{H}(t)) \equiv& \nonumber \\
& \frac{1}{2} \tr{ \left\{  H(0) - \langle H(0) \rangle ,  \widetilde{H}(t) -  \langle\widetilde{H}(t)\rangle  \right\} \rho },&\label{eq:Q_Covariance} 
\end{eqnarray}
with $\{\cdot,\cdot\}$ denoting the anti-commutator.
Since for the chosen initial state $\rho = \ket{+}\!\!\bra{+}$ the mean energy $\langle H(0)\rangle = 0$, the real part of the correlation function coincides with the quantum covariance.

The imaginary part of the quantum correlation function is 
\begin{equation}
\mathrm{Im}\langle \widetilde{H}(t) H(0) \rangle = \frac{1}{2i} \, \tr{\rho \, \left[ \widetilde{H}(t) , H(0) \right]},
\end{equation} 
which is  zero if $[\widetilde{H}(t),H(0)]=0$. For our model, this  occurs at $t =2\pi m/\Omega$ for any integer $m$ (we note that  $\mathrm{Im}\langle \widetilde{H}(t) H(0)\rangle =0$ for $t =\pi/\Omega$, even if $[\widetilde{H}(t),H(0)]\neq 0$).
Conversely, when $\mathrm{Im}\langle \widetilde{H}(t) H(0)\rangle$ is different from zero, none of the operators in the set $\{\widetilde{H}(t),H(0),\rho\}$ commute with each other. This implies that $\mathrm{Im}(q_{if})\neq 0$, and the KDQ is not a joint probability distribution. 

\bigskip


In the following, we provide a physical interpretation of the real and imaginary parts of the quantum correlation function by studying the thermodynamic meaning of the statistical moments of the work distribution. We also compare these results with the ones obtained for a TPM scheme, i.e., for an initial state without energy coherence. 
We further discuss how the KDQ can be used to analyze other thermodynamic properties of a quantum system and its connection to uncertainty relations.

\subsection{Mean work and KDQ work variance}

We first focus to the statistical moments, according to the KDQ, of the work variable 
\begin{equation}
W_{if} \equiv E_f(t) - E_i(0).    
\end{equation}
As the KDQ has the correct marginals~\cite{companion_theory_paper,GherardiniTutorial}, the mean work reads as
\begin{equation}\label{eq:mean_work_kd}
    \langle W \rangle_{\rm KD}  \equiv \sum_{if} q_{if} W_{if} =  \tr{H(t)\rho(t)} - \tr{H(0)\rho}, 
\end{equation} 
which coincides with the unperturbed average energy change given by the difference between the average energy at the final and initial times. While looking trivial from a classical perspective, in the presence of incompatible observables and states, a ``classical'' TPM procedure does not recover this result due to the invasiveness of the measurement process.

\begin{figure}
    \centering
	\includegraphics[width=0.9\columnwidth]{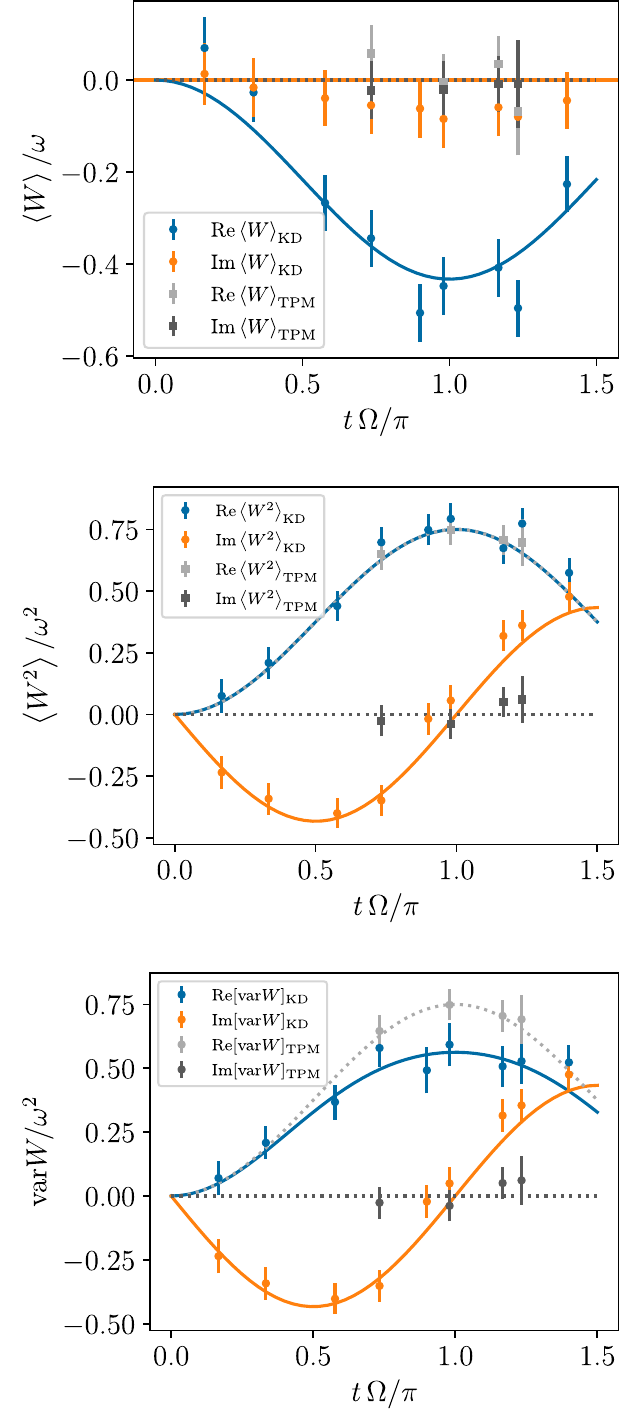}
    \caption{
    Top panel: Mean work computed over the KDQ and TPM distributions. Middle panel: Mean squared work (or 2nd statistical moment) [KDQ (solid lines) and TPM (dotted lines)]. Bottom panel: Work variance [KDQ (solid lines) and TPM (dotted lines)].
    }
    \label{fig:MeanWorkAndVariance}
\end{figure}

In Fig.~\ref{fig:MeanWorkAndVariance}(a) we show $\langle W \rangle_{\rm KD}$ obtained from the experimental values of $q_{if}$ and the simulated unperturbed average energy change~\cite{Baumer2018} $\tr{H(t)\rho(t)} - \tr{H(0)\rho} = -\frac{\delta \Omega}{\omega}\sin^2\left(\frac{\Omega t}{2}\right)$. 
The imaginary part of $\langle W \rangle_{\rm KD}$ is equal to zero, as predicted by theory. In contrast, the real part is always negative (or zero when $\Omega t$ is a multiple of $2\pi$), which means that the quantum system is giving away energy in the form of microwave radiation --- the average work is always extractable work. Such a feature originates from the presence of quantum coherence $\chi$ in the initial state $\rho$, with respect to the eigenbasis of $H(0)$. In fact, the coherence contribution to the average work obtained from the TPM scheme (i.e., $\langle W \rangle_{\rm TPM}$) is always zero, yielding
\begin{equation}
    \langle W \rangle_{\rm KD} - \langle W \rangle_{\rm TPM} = {\rm Tr}\left(U\chi U^{\dagger}H(t)\right).
\end{equation}
In our case study, $\langle W \rangle_{\rm TPM}=0$, 
and we can thus conclude that all the work that can be extracted on average from the system is due to the dynamical processing of the initial quantum coherence --- all the work is quantum work in this context.

When considering the distribution of a stochastic variable, higher (central) moments provide valuable information beyond the mean value, starting from the variance: 
\begin{equation}
    {\rm var}W_{\rm KD} \equiv \langle (W  - \langle W \rangle_{\rm KD})^2\rangle_{\rm KD} = \langle W^2 \rangle_{\rm KD} - \langle W \rangle^2_{\rm KD} \,.
\end{equation}
The behavior of $\langle W^2 \rangle_{\rm KD} \equiv \sum_{i,f} q_{if} W_{if}^2$ as a function of time is shown in Fig.~\ref{fig:MeanWorkAndVariance}(b), while the experimental and simulated values for the work variance are in Fig.~\ref{fig:MeanWorkAndVariance}(c). 
The work variance of the KDQ distribution contains both a real and an imaginary part:
\begin{align}
    {\rm var}W_{\rm KD} = V_R + i V_I .
\end{align}
As in classical statistics, the real part can be decomposed as
\begin{equation}\label{eq:varmh}
    V_R = \var[H(0)] + \var[\widetilde{H}(t)] - 2\,{\rm Cov}\left( H(0),\widetilde{H}(t) \right),
\end{equation}
where the quantum covariance ${\rm Cov}(H(0),\widetilde{H}(t)) \in \mathbb{R}$ has been defined in Eq.~\eqref{eq:Q_Covariance}, and the variance of an operator $A$ is defined as usual as $\var[A] = \tr{\rho  A^2} - \tr{\rho A}^2$.

The imaginary part of the variance is, clearly, a purely quantum term that reads 
\begin{align}\label{eq:V_I}
    V_I &= i \, \tr{\rho \, [H(0),\widetilde{H}(t)]} \\
    &= 2\Im \; \langle H(0)  \widetilde{H}(t)\rangle \,. 
    \label{eq:variance_Im}
\end{align}
This means that $V_I$ is twice the imaginary part of the quantum correlation function $\langle \widetilde{H}(t) H(0) \rangle$ and directly quantifies the non-commutativity between the initial and final energy observables brought about by the quantum work protocol.

Overall, we have seen that the first two central moments of the KDQ encode relevant information about energy fluctuations in a fully \emph{quantum} regime, and we have discussed how non-commutativity affects them.

\subsubsection*{Comparison with the TPM scheme}

To compare our results with the ones obtained in the limit of commuting operators, let us consider the initial state to be diagonal in the initial energy basis and given by $\rho_{\rm d} = \sum_{i} p_i \Pi_i(0)$. Being $p_i = \tr{\rho \Pi_i(0)}$,  $\rho_{\rm d}$ is the completely dephased version, in the initial energy eigenbasis, of the initial state $\rho$ considered so far. This ensures that $[\rho_{\rm d},H(0)]=0$, and thus that the KDQ coincides with the joint probability distribution obtained from the TPM protocol.

For our experiments, given $\rho = \ket{+}\!\!\bra{+}$ that is the state with the maximum coherence with respect to $H(0)$, we have $\rho_{\rm d} = \mathbb{I}/2$. Therefore, in order to obtain the TPM statistics, we repeated the measurements described in the previous section half of the times for the state $\ket{+}\!\!\bra{+}$ and half for the state $\ket{-}\!\!\bra{-}$. This corresponds to considering an equal mixture of the two states, and thus to take $\rho_{\rm d} = \mathbb{I}/2$ as the initial state.

The dataset for the initial state $\rho_{\rm d}$ was analyzed using the same procedure as for the full KDQ. 
The results for the mean work, mean squared work, and work variance are shown in Fig.~\ref{fig:MeanWorkAndVariance}. 
As expected, the imaginary part of all these quantities is zero; moreover, $\langle W \rangle_{\rm TPM} = 0$, and $\langle W^2 \rangle_{\rm TPM} = \Re \langle W^2 \rangle_{\rm KD}$. 
Accordingly, the variance ${\rm var}W_{\rm TPM} > V_R = \Re {\rm var}W_{\rm KD}$. Therefore, the system is performing more work (giving away more energy) and with a smaller variance when the work statistics are computed with respect to the initial state $\rho$, compared to its dephased counterpart, the non-coherent state $\rho_{\rm d}$.

The fact that $\langle W\rangle_{\rm TPM} = 0$ while $\langle W^2 \rangle_{\rm TPM} \neq 0$ shows that there is an energy exchange occurring during the dynamics. It just so happens that the classical mixture of the initial state is such that, on average, the system is not losing or winning energy. 
This is not the case when the initial state is the coherent superposition $\rho$, whereby $\langle W \rangle_{\rm KD} \neq 0$.

\subsection{Relation with the operator of work}

Here we focus on the concept of \textit{operator of work}~\cite{PerarnauLlobetPRL2017,Allahverdyan2005}, defined as $\mathcal{W} = \widetilde{H}(t) - H(0)$, where $\widetilde{H}(t) = U^\dagger H(t) U$ is the final Hamiltonian evolved according to the Heisenberg picture. It is worth mentioning that, even by introducing the operator of work $\mathcal{W}$, we do not end up with a comprehensive definition of quantum work, as the physical meaning of the third and higher moments is not clear so far~\cite{Allahverdyan2005}. 
In conformity with the no-go theorem in Ref.~\cite{companion_theory_paper}, such lack of a complete description of the quantum work is due to the fact that, in general, $[\widetilde{H}(t),H(0)]\neq 0$ and $[\rho,H(0)]\neq 0$, with $\rho$ the initial density operator at the beginning of a given thermodynamic transformation. 
However, notably, the first and second moments of $\mathcal{W}$, computed with respect to $\rho$, have physical interpretations in connection with the KDQ. In this regard, we recover that, in the commutative limit, $\tr{\mathcal{W}\rho}$ and $\tr{\mathcal{W}^2\rho}$ have a clear classical correspondence. Formally, one can show that 
\begin{align}
    \tr{\mathcal{W}\rho} &= \langle W \rangle_{\rm KD} \\
    \tr{\mathcal{W}^2\rho} & = \mathrm{Re} \langle W^2 \rangle_{\rm KD} \\
    {\rm var}\mathcal{W} &= \Re[{\rm var}W_{\rm KD}] ,
\end{align}
where ${\rm var}\mathcal{W} \equiv \tr{\rho (\mathcal{W} - \tr{\mathcal{W}\rho})^2}$ is the so-called \emph{work dispersion}~\cite{Allahverdyan2005}. Therefore, the real part of the KDQ variance of work has the same physical meaning as the work dispersion in \cite{Allahverdyan2005}, which is an already accepted generalization of the variance of work in the quantum case.

In general, $\tr{\mathcal{W}^m\rho} \neq \langle W^m \rangle_{\rm KD}$ (nor to its real part), for $m>2$. These two quantities are equal to each other in the fully commutative case $[\widetilde{H}(t),H(0)] = [\rho,H(0)] = 0$. This remark is pertinent because a primary concern with the work operator $\mathcal{W}$ is that its higher moments lack physical meaning. In fact, for $m>2$, $\tr{\mathcal{W}^m\rho}$ do not coincide with $\langle W^m \rangle_{\rm TPM}$, even when $[\rho,H(0)] = 0$. 
In contrast, we recall that the condition $[\rho,H(0)] = 0$ implies that the KDQ is actually a joint probability, as returned by the TPM scheme. Hence, under such condition, the physical interpretation of $\langle W^m \rangle_{\rm KD}$ coincides with the one of $\langle W^m \rangle_{\rm TPM}$ for any $m$.

\subsection{Connection with the uncertainty relations}

The uncertainty principle is one of the well-known quintessential properties of quantum mechanics. In particular, the Robertson-Schr{\"o}dinger uncertainty relation (RSUR)~\cite{RobertsonPR1929,SchrodingerPmK1930,schrodinger1999heisenberg} takes into account observable correlation to provide a tighter bound. For our observables of interest (the initial and final energy), the RSUR is  
\begin{eqnarray}\label{eq:Heisenberg_Rob_inequality}
\var[H(0)]\var[\widetilde{H}(t)] &\geq& {\rm Cov}(H(0),\widetilde{H}(t))^2 + \nonumber \\
&+& \left(\mathrm{Im}\langle \widetilde{H}(t) H(0) \rangle\right)^{2}.
\end{eqnarray}
We refer the reader to the Supplementary Note 3 for the formal derivation of this result. Notice that the second term on the right-hand side of \eqref{eq:Heisenberg_Rob_inequality} is related to the KDQ work variance, $V_I^2/4$ (see Eq.~\eqref{eq:variance_Im}). Thus, the interferometric reconstruction of the KDQ gives us access to the commutator of the energy observables and, in turn, allows us to characterize the uncertainty in the process under scrutiny.

\begin{figure}
\centering
\includegraphics[width=0.9\columnwidth]{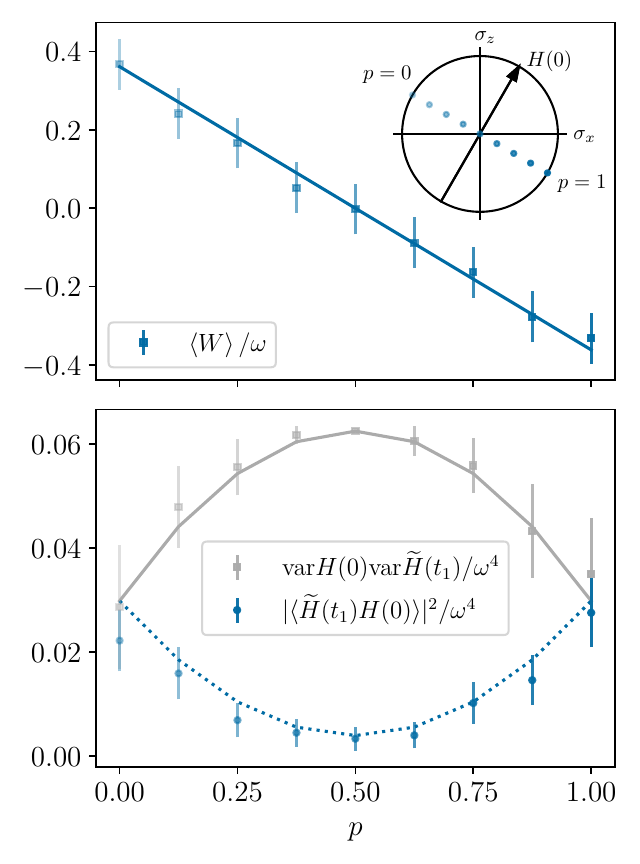}
\caption{
Robertson-Schr\"odinger uncertainty relation (RSUR) [see Eq.~\eqref{eq:Heisenberg_Rob_inequality}] for different initial states. The x-axis, $p$, is a parameter of the initial state $\rho_p = p \ket{+}\!\!\bra{+} + (1-p) \ket{-}\!\!\bra{-}$. Here, we show the data only for one value of the final time $t = (11/15) \Omega/\pi$. 
Top panel: Normalized mean work. As we change the initial state, the mean work changes, going from positive (work done on the system) to negative (work done by the system).
Bottom panel: Normalized left-hand-side of the RSUR, i.e., the uncertainty of $\widetilde{H}$ (gray squares -- experiment; gray line -- theory) and normalized right-and-side of the RSUR, i.e., the lower bound of the uncertainty (blue circles -- experiment; blue dotted line -- theory). 
The light-to-dark tone of the makers, proportional to $p$, indicates the initial state as shown in the inset.}
\label{fig:RS_uncertainty}
\end{figure}

The RSUR [Eq.~\eqref{eq:Heisenberg_Rob_inequality}] gives a lower bound to the uncertainty associated with the observables $H(0)$ and $\widetilde{H}(t)$. 
Fig.~\ref{fig:RS_uncertainty}  shows how the two sides of the RSUR vary when changing the initial state in our experiment. In particular, we assume that the initial state is $\rho_p = p \ket{+}\!\!\bra{+} + (1-p) \ket{-}\!\!\bra{-}$; see the inset in Fig.~\ref{fig:RS_uncertainty}(a). For any value of $p$, these initial states have vanishing mean initial energy. Thus, the right-hand-side of the RSUR is $\big|\langle \widetilde{H}(t) H(0) \rangle\big|^{2}$ that is the quantity that we experimentally reconstruct. 
It is interesting to note that the RSUR bound is saturated when the initial state is pure. This is because we are considering two-level systems as highlighted in~\cite{VegaEJP2021}. 

\section{Discussion}

Quasiprobabilities have been shown to play a crucial role in understanding the effects of incompatible observables on work quantum processes ~\cite{levy2020quasiprobability,lostaglio2018quantum,hernandez2022experimental,GherardiniTutorial}. However, they are generally not directly accessible via projective measurement schemes~\cite{PerarnauLlobetPRL2017}, as one would expect for quantum correlation functions between incompatible observables.

In this paper, we demonstrate that it is possible to experimentally reconstruct the real and imaginary parts of the Kirkwood-Dirac quasiprobability (KDQ) distribution of work. We achieve this by implementing an interferometric scheme, where the characteristic function is mapped onto the coherence of an auxiliary qubit coupled to the system. The characteristic function, encoded into the auxiliary system, is then measured with a Ramsey scheme. 
This interferometric scheme was originally proposed to access the work characteristic function associated with an initial thermal state. In that case, the non-commutativity of quantum mechanics played no role. Here, instead, we exploit the same scheme for arbitrary initial states to reconstruct the full KDQ of work and explore the role of coherences and non-commutativity.

As an experimental platform, we used the electronic spin of a single NV center in diamond as a qubit (system) and its nitrogen nuclear spin as the auxiliary qubit. Our method can be directly implemented in any two-qubit system where the interaction Hamiltonian is proportional to $\sigma_z\sigma_z$. 
One of the main advantages of using an NV center and its nitrogen nuclear spin for this experiment, compared to previous implementations of similar interferometric schemes~\cite{BatalhaoPRL14,PalPRA19}, is that in our platform we can optically initialize both spins into a pure state, and thus prepare initial states that are either mixed or pure.

A common criticism of quasiprobabilities in work processes is their lack of physical interpretation compared to the joint probability distribution of work. 
Here we took a step to solve this issue by giving a physical interpretation of the first and second moments of work associated with the KDQ. We first note that the KDQ has a simple physical interpretation in the commutative case since it coincides with the work joint probability distribution for an initial state that commutes with the initial Hamiltonian~\cite{companion_theory_paper,GherardiniTutorial}. This positions the KDQ as a promising distribution for understanding work statistics.

In agreement with previous results~\cite{hernandez2022experimental}, here we show that the first moment of work coincides with the unperturbed work. While the physical interpretation of higher moments of work is typically lacking, in this paper we provide an interpretation for the work variance, which we reconstructed experimentally with our interferometric scheme. The real part of the work variance follows the same interpretation as its classical limit [see Eqs.~\eqref{eq:varmh}], given in terms of the variance of the initial and final energy and their (quantum) co-variance.  Moreover, we observed that, for an initial state $\rho$ with coherences in the initial Hamiltonian basis, the variance is smaller than the one of a mixed state that corresponds to the dephased version of this initial state. This property goes together with the evidence that the mean work is negative for the initial coherent state but is zero for its corresponding dephased initial state. Therefore, keeping the coherences of the initial state not only increases the extractable work from the system but also makes its variance smaller.

The imaginary part of the work variance has no classical counterpart, as expected. However, it can be interpreted as a measure of the non-commutativity between the initial state and the observables (initial and final Hamiltonian). In this sense, it is a witness of non-classicality. In addition, the imaginary part is relevant for the bounds of the uncertainty between the incompatible observables described by the Robertson-Schr{\"o}dinger inequality. Based on this inequality, we show that the presence of the imaginary part of the KDQ work variance yields a tighter bound for its real part than the bound in the case of a dephased initial state. Finally, we show that the first and second moments of the KDQ work distribution coincide with the first and second moments of the so-called operator of work~\cite{Allahverdyan2005}.

Our results show that it is possible to experimentally reconstruct the full KDQ for closed system dynamics. The same protocol should allow to reconstruct the energy variation statistics for open system dynamics. This paves the way to experimentally investigate energy fluctuations in processes where work and heat are present, either in different strokes~\cite{SolinasPRAmeasurement} or contemporaneously~\cite{HernandezGomez21}. This is very relevant for the characterization of fluctuations in heat engines, e.g.~an Otto-cycle heat engine, where the system is out of equilibrium at the end of the stroke. 
Moreover, demonstrating the possibility of measuring the KDQ in a single NV setup is the first step towards measuring these quasiprobabilities in more complex systems, e.g.~formed by more spins as in the case of ensembles of NV centers, where many-body physics becomes relevant. 
Finally, this interferometric scheme can be easily modified to include more couplings with the auxiliary system, which would enable reconstructing generalized KDQ that are closely related to out-of-time-ordered-correlators~\cite{yunger2018quasiprobability,alonso2019out}. 

\section{Methods}

Here we provide further details on the experimental implementation of the circuit in Fig.~\ref{fig:NV_lattice_and_E_levels_scheme_experiment}c. Specifically, we describe the different steps: Initialization and readout, nuclear spin gates, electronic spin gates, and conditional unitary gates.

\textit{Initialization and readout:}
One of the great advantages of NV centers is that they can be optically initialized and read out. Although the nuclear spins are not directly affected by the laser excitation, their coupling to the NV electronic spin perturbs the nuclear spin state during optical illumination. In our experiments, we thus chose an external bias field  $B_z$ that achieves a good compromise between the two competing tasks of nuclear spin polarization, and electronic spin readout independent of the nuclear spin state. 
Indeed, when the bias field $B_z$ is relatively close to the excited state level anti-crossing (ESLAC)\cite{Jacques09} a $\sim 20$~$\mu$s  green laser illumination pumps the nuclear spin into the $m_I=+1$ state. Therefore, at the beginning of each experimental cycle a long laser pulse ($\sim 20$~$\mu$s) is applied, ensuring that the initial state is always $\left|m_S,m_I\right\rangle = \left|0,+1\right\rangle$.

The same mechanism that initializes the NV spin implies that the average photoluminescence~(PL) intensity is different for the $m_S=0,-1$ levels. Therefore, we can readout the population of the electronic spin by measuring the average PL and normalize with respect to the reference levels for $m_S=0,-1$. 
Note that the bias field is chosen far enough from the ESLAC so that the PL during the first few hundreds of nanoseconds is independent of the nuclear spin and spin-spin correlations. 
In our protocol, we wish to measure the population on the nuclear spin. To achieve this, we first apply a short laser pulse ($\sim400$~ns) that re-initializes the electronic spin into $m_S=0$, but leaves the nuclear spin almost unperturbed. Then we apply a CNOT gate mapping the nuclear spin state onto the electronic spin, and we measure the population of the electronic spin.

\textit{Nuclear spin gates: }
As explained in the main text, the nuclear spin control always satisfies the condition $\Omega_n\ll A$, thus we always  drive the transition $\ket{0,0}\leftrightarrow\ket{0,+1}$, without affecting the $\ket{-1,m_I}$ states, that is, all nuclear gates are conditional (selective.)
A single drive is enough for the first $\pi/2$-pulse shown in Fig.~\ref{fig:NV_lattice_and_E_levels_scheme_experiment}c, since the electronic spin is in the $m_S=0$ state. 
However to implement the (non-selective) final $\pi/2$-pulse, we  need to apply two selective $\pi/2$-pulses with an intermediate non-selective $\pi$-pulse on the electronic spin. 
We note that during the nuclear selective pulses, the $\ket{-1,m_I}$ manifold does evolve under the hyperfine term, $A\sigma_z^{(n)}$. Hence we calibrate the RF $\pi/2$-pulses to have a pulse length that is an even multiple of $2\pi/A$, canceling out the effect of the hyperfine term.

\textit{Electronic spin gates: }
We use on-resonance ($2\pi\nu = \omega_{e}$) high power microwave  ($\Omega\gg A$) to drive the electronic spin non-selectively. The limiting factor in our experiment is the rate at which we can turn on and off the microwave pulses. For this reason, the maximum Rabi frequency we can get is $\Omega/2\pi \simeq  25 $~MHz. This control is a fast enough to neglect the hyperfine term in the Hamiltonian during pulses with duration on the order of $1/\Omega$.

\textit{Conditional unitary gates: }
We implement the conditional unitary gates shown in Fig.~\ref{fig:scheme_original} by exploiting the hyperfine term $H_I$. We can achieve this by decomposing $G_1(u)$ and $G_B(u)$ as:

\begin{align}
    G_1(u) &\overset{(*)}{=} R^{(e)}_{y}(\theta)\, R^{(n)}_{z}(\tfrac{uA}{2})\,  {\rm e}^{-i u H_{\rm I}}  R^{(e)}_{y}(-\theta) \\
    G_B(u) &\overset{(*)}{=} R^{(e)}_{y}(\theta_2)\, R^{(e)}_{z}(-uA) R^{(n)}_{z}(\tfrac{uA}{2})  {\rm e}^{-i u H_{\rm I}}  R^{(e)}_{y}(-\theta_2) ,
\end{align}
where $\overset{(*)}{=}$ indicates equal operators up to a global phase, $R_{w}^{(n)}(\phi) \equiv \mathbb{I} \otimes {\rm e}^{-i \phi \sigma_w}$, $R_{w}^{(e)}(\phi) \equiv {\rm e}^{-i \phi \sigma_w}\otimes \mathbb{I}$, $\theta = \arctan\Omega/\delta$, and $\theta_2 = \theta + \pi$.
We recall that the local gates on the electronic spin are short enough so that we can ignore the effect of the hyperfine term. Moreover, the last two gates in $G_B$ can be ignored because they act on the electronic spin and will not affect the measurement of the nuclear spin population. Finally, we can avoid implementing the local gates $R^{(n)}_{z}(\tfrac{uA}{2})$ by accordingly changing the phase of the second (and third) $\pi/2$ pulses for the nuclear spin. 
The result is the pulse sequence depicted in Fig.~\ref{fig:NV_lattice_and_E_levels_scheme_experiment}c.

\section*{DATA AVAILABILITY STATEMENT}
The data that support the findings of this study are available from the corresponding author upon reasonable request.

\begin{acknowledgments}
We thank Matteo Lostaglio for several discussions involving this work and for carefully reading the manuscript. We are grateful to Guoqing Wang for insightful discussions. We wish to acknowledge financial support from the MISTI Global Seed Funds MIT-FVG Collaboration Grants ``Non-Equilibrium Thermodynamics of Dissipative Quantum Systems (NETDQS)'' and ``Revealing and exploiting quantumness via quasiprobabilities: from quantum thermodynamics to quantum sensing'', the project PRIN 2022 Quantum Reservoir Computing (QuReCo), and the PNRR MUR project PE0000023-NQSTI funded by the European Union--Next Generation EU. 
The work was also supported by the European Commission under Grant No. 101070546– MUQUABIS, and by the European Union's Next Generation EU Programme through the I-PHOQS Infrastructure, and through the Project PRIN 2022 QUASAR. 
A.~Belenchia acknowledges support
from the Horizon Europe EIC Pathfinder project
QuCoM (Grant Agreement No. 10104697). This work was in part supported by the Center for Ultracold Atoms (an NSF Physics Frontiers Center), PHY-2317134. A.~Levy acknowledges support from the Israel Science Foundation (Grant No.~1364/21). T.~Isogawa acknowledges support from the Keio University Global Fellowship.
\end{acknowledgments}

\section*{Author Contributions}
S.H.G., A.B., A.L., N.F., S.G. conceived the project. S.H.G. and T.I. designed and performed the experiments. T.I. designed the pulse sequence. S.H.G. carried out the analysis. P.C. supervised the project. All authors contributed to the interpretation of the results and to writing the manuscript.

\section*{Competing Interests}
The Authors declare no Competing Financial or Non-Financial Interests.

\bibliography{KDQ_interferometric_scheme}

\newpage
\onecolumngrid
\setcounter{section}{0}
\setcounter{subsection}{0}

\section*{Supplementary information}

\subsection*{Supplementary Note 1: Properties of the unitary evolution and interferometric scheme}

As we argue in the main text, the form of the time-dependent Hamiltonian [Eq.~(8) in the main text] 
allows us to write the unitary operator describing the whole system dynamics as
\begin{equation}\label{cvd}
U(t) = \exp\left(-i  t \delta \sigma_z /2 \right)\exp\left(-i  t \Omega \sigma_x /2 \right) \otimes \mathbb{I}. 
\end{equation}
We can see this explicitly as follows. Let's start by performing a change of frame via the unitary 
$V(t)=\exp\left(i  t \delta \sigma_z /2 \right)$ such that the Hamiltonian in the new reference frame is
\begin{equation}
    \tilde{H}(t)=V(t)H(t)V^\dag(t)+i\frac{dV(t)}{dt}V^\dag (t) \,.
\end{equation}
The last term in the previous equation is simply $\delta\sigma_z/2$. For the other terms, recalling that $V(t)=\cos\left(t \delta /2 \right)\mathbb{I}+i\sin\left(t \delta /2 \right) \sigma_z $, and using the Pauli algebra we can easily obtain
\begin{align}
    & V(t)\,\sigma_x V^{\dag}(t)=\cos(t\delta)\sigma_x-\sin(t\delta)\sigma_y\label{mathex}\\
    & V(t)\,\sigma_y V^{\dag}(t)=\cos(t\delta)\sigma_y+\sin(t\delta)\sigma_x.\label{mathex2}
\end{align}

Putting this all together, we have that $\tilde{H}(t)=\Omega\sigma_x/2$. In this frame, a state evolves as $\ket{\tilde{\psi}(t)}=\exp{-i\tilde{H}t}\ket{\psi(0)}$. Moving back to the original frame we see that, since $\ket{\tilde{\psi}(t)}=V(t)\ket{\psi(t)}$, the evolution operator is indeed given by Eq.~\eqref{cvd}.

These same manipulations can be used to see the equivalence between the interferometric scheme involving $G_2$ and $U(t)$ and the one with $G_B$ and $U_B$ as discussed in the main text. Indeed, in order for the output of the quantum circuit in Fig.~\ref{fig:scheme_original}, 
with $G_B$ and $U_B$ replacing $G_2$, to coincide with the real and imaginary part of the characteristic function for the KDQ, we need
\begin{equation}
    e^{i u H_t}=e^{-i t \delta \sigma_z/2}e^{i u H_0}e^{i t \delta \sigma_z/2}.
\end{equation}
Let us consider the right-hand side of this equation. We have $e^{-i t \delta \sigma_z/2}e^{i u H_0}e^{i t \delta \sigma_z/2}$. Using then Eqs.~\eqref{mathex} and~\eqref{mathex2} (with $\delta\rightarrow -\delta$) and the fact that $H_0=\Omega\sigma_x/2+\delta\sigma_z/2$ we conclude that
\begin{equation}
    e^{-i t \frac{\delta}{2} \sigma_z} H_0 \, e^{i t \frac{\delta}{2} \sigma_z}=\frac{\Omega}{2}\left(\cos(\delta t)\sigma_x+\sin(\delta t)\sigma_y\right)+\frac{\delta}{2}\sigma_z.
\end{equation}

\subsection*{Supplementary Note 2: Additional details for the comparisson between theory and experiment}

As described in the main text, the reconstruction of the real and imaginary parts of the KDQ $q_{if}$ is achieved by measuring the characteristic function $\mathcal{G}(u)$, obtaining the Fourier transform to reconstruct $P(W)$, and then integrating in the vicinity of the allowed values of $W$, namely at $W=0,\pm \omega/2$. 
In this section we explore two possible sources of error during this protocol, and how they can lead to discrepancies between experiment and theory. 

\subsubsection{Self consistency test}
First, we analyze the effect of the Fourier analysis due to a finite number of experimental data points when measuring $\mathcal{G}(u)$. 
On the one hand, the maximum time $u$ that can be used during the experiments is limited by the finite coherence time of the spin qubit $T_2^*$. On the other hand, reducing the time step between consecutive values of the time $u$ is limited by the precision of the mw pulses ($\sim 10$~ns). Due to this, even in the ideal case of an experiment without noise (simulation), the Fourier analysis of the data will result in some small discrepancy between the original and the reconstructed $q_{if}$. 
This is shown by the solid and dashed lines in Fig.~\ref{fig:KDQ_variation_offset_contrast}. 
The solid lines (same as in Fig.~5 in main text) represent the values of the KDQ simulated as $q_{if} = \tr{ U^\dag \Pi_f U \; \Pi_i \, \rho }$, see Eq.~(2) in the main text. 
The dashed lines are obtained as follows: First, we consider the same values of $q_{if}$ as for the solid line, but we use them to obtain the characteristic function $\mathcal{G}_{\mathrm{sim}}(u) = \sum_{i,f}q_{if}(\rho)\; e^{ - iu (E_f -E_i)} $ for a set of values for the gate time $u$ that are comparable with the experimental ones. Then, the Fourier transform is applied to recover $P_{\mathrm{sim}}(W)$ and, finally, we integrate in the vicinity of $W=0,\pm\omega/2$ to recover $q_{if}$. 
It is clear from the figure that the solid lines and the dashed ones are not the same. This discrepancy coming from a finite sampling of data points is comparable to the experimental precision. In Fig.~\ref{fig:KDQ_variation_offset_contrast} we also show the experimental data (same as in Fig.~5 in the main text). The experimental data is closer to the dashed line than to the solid line because also the experimental data was reconstructed from the Fourier analysis of the characteristic function. 

Up to this point, we only considered an ideal simulation of the characteristic function. Now we investigate the effect of noise in the simulation. 

\subsubsection{Offset and amplitude noise}
\begin{figure*}
    \centering
    \includegraphics[width=\textwidth]{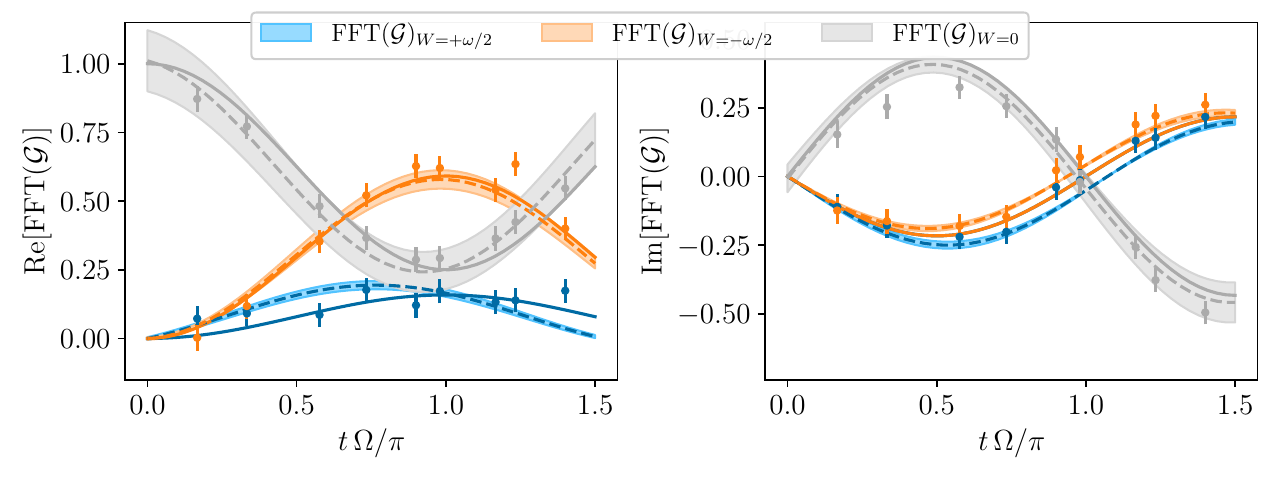}
    \caption{
    Reconstruction of the real and imaginary values of the KDQ $q_{if}$. 
    The solid lines and the bullets with error bars are exactly the same as in Fig.~5 of the main text. The dotted line is the result of an ideal experiment without noise (see text).  
    The shaded area represents the effect of adding some random offset or a random amplitude variation to the ideal experiment. .}
    \label{fig:KDQ_variation_offset_contrast}
\end{figure*}

We recall that the outputs of the experiment, $\langle\sigma_x\rangle = \mathrm{Re}\mathcal{G}(u)$ and $\langle\sigma_y\rangle = \mathrm{Im}\mathcal{G}(u)$, are the result of measuring the intensity of the red light emitted by the NV center and comparing it with the previously characterized intensity of the eigenstates of the observable. 
To be more clear, the value $\langle\sigma_x\rangle$ is obtained as $2(s-s_+)/(s_- -s_+)-1$, where $s$ is the photoluminescence (PL) intensity measured after running the full experiment, as described in the main text, and $s_{+(-)}$ is the PL intensity measured in a different experiment where the spin qubit is simply prepared in the eigenstate $\ket{+}(\ket{-})$ of the observable $\sigma_x$. An equivalent protocol is used to measure $\langle\sigma_y\rangle$. 
The variations on the PL intensity for the signal $s$ are considered in the error bars of the experimental data, which are the standard deviation over the $\sim 10^6$ repetitions of each single experiment. 
However, some small error in the microwave or radiofrequency pulses may lead to a small variation of $s_{+-}$ that produces variations in the offset and/or contrast (amplitude) of the measured signal -- $\mathcal{G}(u)$. These variations are small enough to be ignored when the experiments are performed, but they result in non-negligible variations after processing the data. 
As an example, we use the same simulated characteristic function $\mathcal{G}_{\mathrm{sim}}(u)$ as before, and we randomly scale its amplitude (from $95\%$ to $105\%$) and we randomly add a small  ($<5\%$) offset. 
These variations are within the error bar of the normalizing signals $s_{+,-}$, and are difficult to recognize by just seeing the data. The simulated values of $\mathcal{G}_{\mathrm{sim}}(u)$ with noise are then analyzed, as before, to reconstruct the values of the KDQ. 
The shaded areas in figure~\ref{fig:KDQ_variation_offset_contrast} represent the region covered by the ideal simulation (dashed line) when a small variation of the characteristic function is introduced. This \textit{realistic} model of the experiment coincides with $91\%$ of the experimental points (within one standard deviation). 
Notice also that the case for $W=0$ (i.e. $q_{00}+q_{11}$) is the most sensitive one to these variations of the characteristic function, as evidenced by the large gray area compared to the blue and orange ones. The reason for this is that the $W=0$ peak for $P(W)$ corresponds to the ``\textit{zero frequency}'' peak in terms of Fourier analysis, and is therefore directly affected by the offset of the signal.

To conclude this section, the experimental data clearly follows the simulated $q_{if}$ (solid line). However, to quantify the agreement between experimental data and simulation, we note that that $57\%$ of the experimental points coincide with the simulation (solid line) up to one standard deviation. And $91\%$ of the experimental points are within the simulation by two standard deviations (solid line). 
While it's not a perfect match, it's important to emphasize that there are no free parameters involved in comparing these two. 
Moreover, the difference between them can be explained by considering the effect of a finite sampling of experimental data and additional noise from the normalization of the data.

\subsection*{Supplementary Note 3: Restrictions to the variance of energy variations due to the uncertainty principle: Formal derivation}

We start from the computation of the energy-variation variance in terms of the KDQ distribution, i.e., $\var[\Delta E]_{\rm KD}$, which is formally defined by
\begin{eqnarray}
    \var[\Delta E]_{\rm KD} &\equiv& \langle\Delta E^2\rangle_{\rm KD} - \langle\Delta E\rangle^{2}_{\rm KD} 
    = \displaystyle{ \sum_{i,f}p_{i,f}\left(\tilde{E}_{f}-E_{i}\right)^2 - \Big(\sum_{i,f}p_{i,f}(\tilde{E}_{f}-E_{i})\Big)^2 }, 
\end{eqnarray}
and simplifies as
\begin{equation}
    \var[\Delta E]_{\rm KD}=\var[H(0)] + \var[\tilde{H}(t)]- \displaystyle{ 2\sum_{i,f}p_{i,f}E_{i}\tilde{E}_{f} + 2\langle H(0)\rangle\!\langle \tilde{H}(t)\rangle}\,,
\end{equation}
where 
\begin{eqnarray}
&&\var[H(0)] \equiv \sum_{i,f}p_{i,f}E_{i}^2 - \langle H(0)\rangle^2 \in \mathbb{R}, \\ 
&&\var[\tilde{H}(t)] \equiv \sum_{i,f}p_{i,f}\tilde{E}_{f}^2 - \langle \tilde{H}(t)\rangle^2 \in \mathbb{R}, \\
&&\sum_{i,f}p_{i,f}E_{i}\tilde{E}_{f} = \tr{ \rho \, H(0) \tilde{H}(t)},
\end{eqnarray}
with $\displaystyle{\langle H(0)\rangle \equiv \sum_{i,f}p_{i,f}E_{i}}$, and $\displaystyle{\langle\tilde{H}(t)\rangle \equiv \sum_{i,f}p_{i,f}\tilde{E}_{f}}$.
The Hamiltonian operators $H(0)$ and $\tilde{H}(t)$ are Hermitian. Then, by observing that 
\begin{equation}
    \tr{ \rho \Big(H(0)-\langle H(0)\rangle\Big) \Big(\tilde{H}(t)-\langle \tilde{H}(t)\rangle\Big)} = \tr{ \rho H(0) \tilde{H}(t)} - \langle H(0)\rangle\!\langle\tilde{H}(t)\rangle,
\end{equation}
we get
\begin{equation}
    \var[\Delta E]_{\rm KD} = \var[H(0)] + \var[\tilde{H}(t)] 
    - 2\tr{ \rho \Big(H(0) - \langle H(0)\rangle\Big)\Big( \tilde{H}(t) - \langle \tilde{H}(t)\rangle \Big) }.
\end{equation}
In general, ${\rm Tr}\Big( \rho \big(H(0)-\langle H(0)\rangle\big) \big(\tilde{H}(t)-\langle \tilde{H}(t)\rangle\big) \Big)$, from now on denoted as ${\rm Tr}\big( \rho \Delta H(0) \Delta\tilde{H}(t) \big)$ with $\Delta H(0) \equiv H(0) - \langle H(0)\rangle$ and $\Delta\tilde{H}(t) = \tilde{H}(t) - \langle \tilde{H}(t)\rangle$, is a complex number whose real and imaginary parts can be determined through the following relation:
\begin{equation}
    \tr{\rho \Delta H(0) \Delta\tilde{H}(t)} = \frac{1}{2}\tr{\rho \{\Delta H(0),\Delta\tilde{H}(t)\}} - \frac{1}{2}i\tr{ i\rho [\Delta H(0), \Delta\tilde{H}(t)]}
\end{equation}
with $\{\cdot,\cdot\}$ and $[\cdot,\cdot]$ denoting the anti-commutator and commutator, respectively. By definition, the quantity $\frac{1}{2}{\rm Tr}\big( \rho \{\Delta H(0), \Delta\tilde{H}(t)\} \big)$ is the \textit{quantum covariance} of the operators $H(0)$ and $\tilde{H}(t)$:
\begin{equation}
    {\rm Cov}(H(0),\tilde{H}(t)) \equiv \frac{1}{2} \tr{\rho \left\{ (H(0) - \langle H(0)\rangle) , (\tilde{H}(t) - \langle \tilde{H}(t)\rangle) \right\}}.
\end{equation}
Accordingly, by noting that 
\begin{equation}
    \tr{ i \rho \, [\Delta H(0), \Delta\tilde{H}(t)] } = \tr{ i \rho \, [H(0), \tilde{H}(t)] },
\end{equation}
one can finally conclude that 
\begin{equation}
    \var[\Delta E]_{\rm KD} = \var[H(0)] + \var[\tilde{H}(t)] -2\,{\rm Cov}(H(0),\tilde{H}(t)) - \tr{ \rho \, [H(0), \tilde{H}(t)] }.
\end{equation}
According to the KDQ distribution, $\var[\Delta E]_{\rm KD}$ is, in general, a complex number and its imaginary part equals to ${\rm Tr}\big( i \rho \, [H(0), \tilde{H}(t)] \big)$, i.e.,
\begin{equation}
    \Im\var[\Delta E]_{\rm KD} = \tr{ i \rho \, [H(0), \tilde{H}(t)] }.
\end{equation}
Secondly, let us define the \emph{correlation matrix}
\begin{eqnarray}
    \mathcal{C} &=& \begin{pmatrix} \tr{ \rho \, \Delta H(0)^2 } & \tr{ \rho \, \Delta H(0) \Delta\tilde{H}(t) } \\
    \tr{ \rho \, \Delta\tilde{H}(t) \Delta H(0) } & \tr{ \rho \, \Delta\tilde{H}(t)^2 }
    \end{pmatrix} =\nonumber\\
    &=&\begin{pmatrix}
    \var[H(0)] & \tr{ \rho \, \Delta H(0) \Delta\tilde{H}(t) } \\
    \tr{ \rho \, \Delta H(0) \Delta\tilde{H}(t) }^{*} & \var[\tilde{H}(t)]
    \end{pmatrix},\nonumber\\
    &&
\end{eqnarray}
where the complex number ${\rm Tr}\big( \rho \, \Delta\tilde{H}(t) \Delta H(0) \big)$ is the conjugate of ${\rm Tr}\big( \rho \, \Delta H(0) \Delta\tilde{H}(t) \big) = {\rm Cov}(H(0),\tilde{H}(t)) + \frac{1}{2}{\rm Tr}\big(\rho \, [H(0),\tilde{H}(t)] \big)$, being $H(0)$ and $\tilde{H}(t)$ Hermitian operators. Therefore, also $\mathcal{C}$ is an Hermitian operator ($\mathcal{C}^{\dag}=\mathcal{C}$). 
Moreover it is positive semi-definite ($\mathcal{C}\geq 0$) which implies that the determinant is nonnegative $({\rm Det[\mathcal{C}]\geq 0)}$,
 and translates to
\begin{equation}
    \var[H(0)]\var[\tilde{H}(t)] \geq {\rm Cov}(H(0),\tilde{H}(t))^{2} + \frac{1}{4}\tr{ i \rho \, [H(0),\tilde{H}(t)] }^{2}.
    \label{eq:DetC}
\end{equation}
This inequality can be directly proven for a pure state $\rho=\ket{\psi}\bra{\psi}$  by defining $\ket{\xi}\equiv \Delta H(0)\ket{\psi}$ and $\ket{\phi}\equiv \Delta \tilde{H}(t)\ket{\psi}$ and noting that ${\rm Det}[\mathcal{C}]= \braket{\xi|\xi}\braket{\phi|\phi}-\braket{\xi|\phi}\braket{\phi|\xi}\geq 0$, where in the last stage, we used the Cauchy–Schwarz inequality. Generalization to mixed states stems from its linearity. Inequality~(\ref{eq:DetC}) can be cast into the form, 
\begin{equation}
    \left|{\rm Cov}(H(0),\tilde{H}(t))\right| \leq \sqrt{\var[H(0)]\var[\tilde{H}(t)] - \dfrac{1}{2}{\rm Tr}\left( i \rho \, [H(0),\tilde{H}(t)]\right)^2 }.
\end{equation}
In this way, by recalling that 
\begin{equation}
\Re \var[\Delta E]_{\rm KD} = \var[H(0)] + \var[\tilde{H}(t)] - 2\,{\rm Cov}(H(0),\tilde{H}(t)),
\end{equation}
and $\Im\var[\Delta E]_{\rm KD} = {\rm Tr}( i \rho \, [H(0), \tilde{H}(t)] )$, 
we obtain the result
\begin{equation}
    \left| \Re \var[\Delta E]_{\rm KD} - \var[H(0)] - \var[\tilde{H}(t)]\right| \leq \sqrt{\var[H(0)]\var[\tilde{H}(t)] - \Im\var[\Delta E]_{\rm KD}^2 },
\end{equation}
that concludes our derivation.

\end{document}